\documentclass[
 aps,
 preprint,
 fleqn
]{revtex4}

\usepackage[T1]{fontenc}
\usepackage{amsmath}
\usepackage{amssymb}
\usepackage{amsfonts}

\usepackage{graphicx}

\usepackage{wick}

\newcommand{\psibar}[1]{\rlap{$\displaystyle \hspace{ #1 ex} \bar{\phantom{\psi}}$}}
\newcommand{\ME}[3]{\langle {#1} | {#2} | {#3} \rangle}

\newcommand{\pslash}{\rlap{$\displaystyle \hspace{-0.4ex} \diagup $}p}
\newcommand{\ppslash}{\rlap{$\displaystyle \hspace{-0.4ex} \diagup $}p'}

\newcommand{\Lslash}{\rlap{$\displaystyle \hspace{-0.2ex} \diagup $}L}
\newcommand{\Deltaslash}{\rlap{$\displaystyle \hspace{-0.1ex} \diagup $}\Delta}
\newcommand{\Gammaslash}{\rlap{$\displaystyle \hspace{-0.3ex} \diagup $}\Gamma}
\DeclareMathOperator{\intd}{d}
\DeclareMathOperator{\e}{e}
\DeclareMathOperator{\T}{T}
\DeclareMathOperator{\Tr}{Tr}

\begin{document}

\title{Evaluation of the Nucleon Helicity Flip Form Factor using One and Two Virtual Photons}
\date{\today}
\author{Thorsten Sachs}
\author{Patrick Sturm}
\affiliation{Institut für Theoretische Physik II, Ruhr-Universität Bochum, D-44780 Bochum, Germany}

\begin{abstract}
In this work, we will evaluate the nucleon helicity flip form factor at the limit of large momentum transfer. Hereby, we will study the exchange of one and two virtual photons separately. For the calculation of the scattering amplitudes and nucleon transition probability matrix elements, we combine QCD perturbation theory with an expansion in nucleon distribution amplitudes. Using the combination of leading and sub-leading twist nucleon distribution amplitudes, one obtains the desired form factor. Using this technique, we will obtain a divergent result for the form factor. Nevertheless, the structure of the divergency can be extracted. Finally, we will comment the obtained expressions and discuss the behavior in unpolarized and polarized cross sections.
\end{abstract}

\maketitle

\section{Introduction}

Elastic electron nucleon scattering mediated by the electromagnetic interaction is the most considered process to receive information about the nucleon structure within QCD. Applying the basic one photon exchange approximation, the required nucleon transition probability matrix elements are traditionally expressed by the Dirac form factor and the Pauli form factor or, equivalently, the magnetic form factor and the electric form factor. For convenience, we use the representation by the magnetic form factor and the Pauli form factor. Moreover, we concentrate our considerations on the proton form factors
\pagebreak[0]
\begin{eqnarray}
\ME{p(P')}{J_\mu^{em}(0)}{p(P)} &=& \bar{N}(P')\left[G_M^p(Q^2)\gamma_\mu-F_2^p(Q^2)\frac{(P'+P)_\mu}{2m_N}\right]N(P) \\
\label{nff}
\ME{n(P')}{J_\mu^{em}(0)}{n(P)} &=& \bar{N}(P')\left[G_M^n(Q^2)\gamma_\mu-F_2^n(Q^2)\frac{(P'+P)_\mu}{2m_N}\right]N(P).
\end{eqnarray}

At large momentum transfer $Q^2=-q^2=-(P'-P)^2$, one just gets the contribution for the magnetic form factor with the power behavior of $Q^{-4}$. This form factor was measured in a comprehensive region and calculated with different techniques, basically with the QCD factorization theorem. Among other form factors, we studied the magnetic form factor in \cite{Sachs:2011b}. For further information, we recommend the references in this work.

Moving to intermediate values of the momentum transfer, one also gets contributions for the Pauli form factor with the power behavior of $Q^{-6}$. The different power behavior arises from the helicity flip of the nucleon and so this form factor is also known as helicity flip form factor. Concerning this form factor, experimental data are also available. Moreover, one has discovered a different behavior depending on the type of the experiment.

The basic information were taken from unpolarized cross sections. Using the Rosenbluth separation technique \cite{Rosenbluth:1950}, several experiments were performed. Hereby, early experiments did not show significant double photon corrections, see \cite{Yount:1962}, \cite{Browman:1965}, \cite{Anderson:1966}, \cite{Bartel:1967}, \cite{Anderson:1968}, \cite{Bouquet:1968}, \cite{Mar:1968}, \cite{Camilleri:1969}, or radiative corrections, see \cite{Tsai:1961}, \cite{Mo:1969}. Further experiments were executed in \cite{Litt:1970}, \cite{Price:1971}, \cite{Bartel:1973}, and with taking into account radiative corrections, see \cite{Walker:1994}, \cite{Andivahis:1994}. Moreover, the available data were fitted in \cite{Bosted:1995}. The consequences of radiative corrections were considered in \cite{Arrington:2004}. The discussed technique is useful at low $Q^2$, but at larger $Q^2$, the contribution of the helicity flip form factor is suppressed by the momentum transfer. However, the electric form factor seems to have the same power behavior as the magnetic form factor. In order to measure the desired form factor at larger values of $Q^2$, one has to study polarized cross sections. During the last years, the experimental requirements have been created and so various experiments have been performed. Hereby, one needs a polarized electron beam. From the experimental perspective, the polarization transfer method, discussed in \cite{Arnold:1981}, seems to be in favor. Hereby, one has to measure the polarization of the recoil proton \cite{Milbrath:1998}, \cite{Jones:2000}, \cite{Gayou:2001}, \cite{Gayou:2002}, \cite{Punjabi:2005}, \cite{MacLachlan:2006}, \cite{Ron:2007}, \cite{Puckett:2010}. The alternative is to use polarized proton targets \cite{Jones:2006}, \cite{Crawford:2007}. Concerning these data, the electric form factor seems to be power suppressed compared to the magnetic form factor. The different measurements were compared in \cite{Arrington:2003}, \cite{Christy:2004}, \cite{Qattan:2005}.

From the theoretical perspective, the calculation of the desired form factor in the large $Q^2$ region is problematic. Using the QCD factorization theorem and the basic one photon exchange, the Pauli form factor was studied in \cite{Belitsky:2003}. In this work, divergent integrals were obtained and therefore a cutoff parameter related to an effective size of the nucleon was introduced. This form factor was also considered in \cite{Brodsky:2004}. Hereby, different models were considered and a cutoff parameter related to an effective mass of the nucleon was discussed with different logarithmic power behavior.

In order to understand the different behavior in the discussed experiments, it has been suggested that the two photon exchange contribution can cause this situation. Therefore, an advanced form factor parametrization was developed in \cite{Guichon:2003}. The modified magnetic form factor was calculated in \cite{Kivel:2009}. The obtained corrections cannot describe the experiments without an input from the helicity flip form factor. Therefore, it is necessary to study the required form factor in one and two photon exchange approximation.

Let us apply the technique specified in \cite{Sachs:2011b} to evaluate the form factor. We have to combine QCD perturbation theory with an expansion in nucleon distribution amplitudes again. Hereby, we have to use the combination of leading and sub-leading twist nucleon distribution amplitudes, studied in \cite{Braun:2000}. We will start with the one photon exchange and we will finish with the two photon exchange. Using our technique, we will obtain a divergent result for the form factor. Nevertheless, the structure of the divergency can be extracted. Concerning the modified helicity flip form factor, we will obtain the dependence on one additional variable. Calculating the experimental cross section of the process and using the momentum transfer and the scattering angle as variables, one gets a different general behavior in the one and two photon exchange approximation. In the first case, the form factor depends on the momentum transfer only. This was already known and so the Rosenbluth separation technique could be applied. In the second case, the form factor depends on the momentum transfer and apart from that, it depends on the scattering angle additionally. That means, the Rosenbluth separation technique cannot be used in this case. Moreover, the obtained power behavior of the helicity flip form factor can describe the experimental data based on polarized cross sections qualitatively. According to this, we can explain the different behavior in the experiments using unpolarized or polarized cross sections. Finally, we will discuss the required modifications to avoid the divergency.

\section{One Photon Exchange Approximation}

Let us start with the presentation of a sample diagram. In the upper part, we see the incoming electron on the left and the outgoing electron on the right. In the lower part, we have the incoming proton on the left and the outgoing proton on the right. The required quark lines denote $u$, $u$, $d$ from top to bottom. The designations at the vertices are the corresponding coordinates and the designations at the lines are the corresponding momenta.

\begin{center}
\includegraphics[scale=1.00]{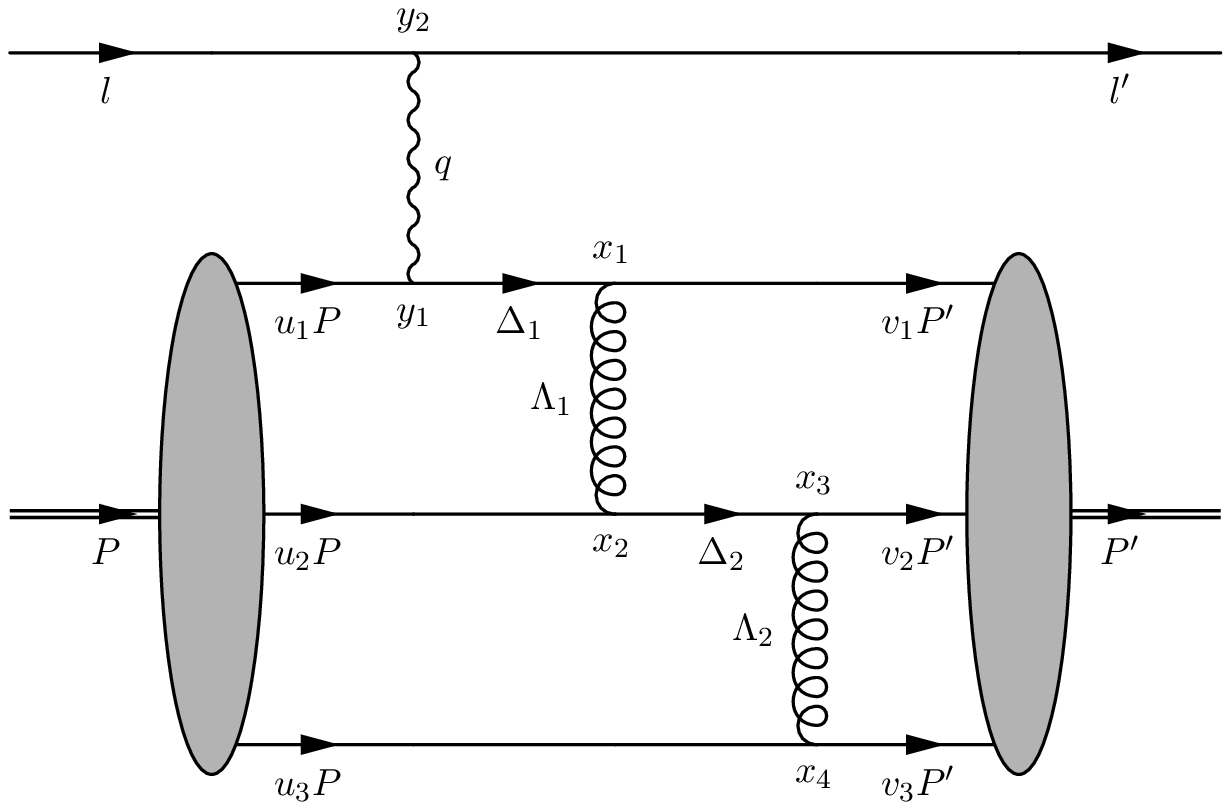}
\end{center}

We need the expression for the scattering amplitude. Applying QED Feynman rules, one can evaluate the leptonic part of the diagram directly
\pagebreak[0]
\begin{equation}
\mathcal{M}=-i(4\pi\alpha_{em})\prod_{i=1}^2\int\frac{\intd^4\!y_i}{(2\pi)^4}\int\frac{\intd^4\!q}{q^2+i0}\bar{u}(l')\gamma^\mu u(l)
\ME{p(P')}{J_\mu^{em}(y_1)}{p(P)}\e^{iq\cdot(y_2-y_1)}\e^{-iy_2\cdot(l-l')}.
\nonumber
\end{equation}

One can replace $\ME{p(P')}{J_\mu^{em}(y_1)}{p(P)}$ with $\ME{p(P')}{J_\mu^{em}(0)}{p(P)}\e^{-iy_1\cdot(P-P')}$. The $y_2$ integration leads to $q=l-l'$ and the $y_1$ integration leads to $q=P'-P$, so that
\pagebreak[0]
\begin{equation}
\mathcal{M}=-\frac{i(4\pi\alpha_{em})}{q^2}\bar{u}(l')\gamma^\mu u(l)\ME{p(P')}{J_\mu^{em}(0)}{p(P)}.
\end{equation}

Using $Q^2=-q^2$ and $2\bar{P}=P+P'$, we introduce the expansion in nucleon electromagnetic form factors
\pagebreak[0]
\begin{equation}
\label{ohf}
\mathcal{M}=\frac{i(4\pi\alpha_{em})}{Q^2}\bar{u}(l')\gamma^\mu u(l)\bar{N}(P')\left[G_M^p(Q^2)\gamma_\mu-F_2^p(Q^2)\frac{\bar{P}_\mu}{m_N}\right]N(P).
\end{equation}

The advantage of this expression is the separation of the leptonic and the hadronic part. Consequently, we only need to consider the matrix element $\ME{p(P')}{J_\mu^{em}(0)}{p(P)}$. This behavior was already used for the calculation of the magnetic form factor, see \cite{Sachs:2011b}.

Applying the S-matrix expansion including the interaction part of the QCD Lagrangian, one gets the following leading expression for the desired matrix element
\pagebreak[0]
\begin{equation}
\frac{(4\pi\bar{\alpha}_s)^2}{24}\ME{p(P')}{\sum_qe_q\bar{\psi}_q(0)\gamma_\mu\psi_q(0)
\T\left[\prod_{i=1}^4\int\intd^4\!x_i\sum_{q_i}\bar{\psi}_{q_i}(x_i)\gamma_{\alpha_i}A^{\alpha_i}(x_i)\psi_{q_i}(x_i)\right]}{p(P)}.
\nonumber
\end{equation}

This expansion can be described by 42 Feynman diagrams and Wick contractions. We can extract the representation of the diagram which we want to study.

Let us begin with the determination of the color factor. Therefore, one has to examine the color structure of the diagram, denoting the color indices with ($a,\ldots,i$). We get
\pagebreak[0]
\begin{equation}
\begin{split}
& \psibar{9.9} \wick[u]{1}{[<1\psi_u(x_1)]_c[>1\psi_u(0)]_a}
\psibar{10.0} \wick[u]{1}{[<1\psi_u(x_3)]_g[>1\psi_u(x_2)]_d}
\wick[u]{1}{[t^{a_1}]_{bc}[t^{a_2}]_{de}<1A_{\alpha_1}^{a_1}(x_1)>1A_{\alpha_2}^{a_2}(x_2)}
\wick[u]{1}{[t^{a_3}]_{fg}[t^{a_4}]_{hi}<1A_{\alpha_3}^{a_3}(x_3)>1A_{\alpha_4}^{a_4}(x_4)} \\
& \ME{p(P')}{[\bar{\psi}_u(x_1)]_b[\bar{\psi}_u(x_3)]_f[\bar{\psi}_d(x_4)]_h}{0}\ME{0}{[\psi_u(0)]_a[\psi_u(x_2)]_e[\psi_d(x_4)]_i}{p(P)}.
\end{split}
\nonumber
\end{equation}
Combining all terms and contracting the generators, one gets the color factor
\pagebreak[0]
\begin{equation}
\mathcal{C}_F = \frac{1}{6}\varepsilon_{bfh}\varepsilon_{aei}\delta_{ca}\delta_{gd}[t^{a_1}]_{bc}[t^{a_2}]_{de}[t^{a_3}]_{fg}[t^{a_4}]_{hi}
\delta^{a_1a_2}\delta^{a_3a_4} = \frac{4}{9}.
\end{equation}

We continue with the evaluation of the Lorentz structure of the diagram, designating the Lorentz indices with ($a,\ldots,j$). Including $\mathcal{C}_F$, we obtain the following expression
\pagebreak[0]
\begin{equation}
\begin{split}
& -\frac{(4\pi\bar{\alpha}_s)^2e_u}{54}\prod_{i=1}^4\int\intd^4\!x_i
[\gamma_\mu]_{ab}[\gamma_{\alpha_1}]_{cd}[\gamma_{\alpha_2}]_{ef}[\gamma_{\alpha_3}]_{gh}[\gamma_{\alpha_4}]_{ij} \\
& \psibar{10.0}
\wick[u]{1}{[<1\psi_u(x_1)]_d[>1\psi_u(0)]_a}
\psibar{10.1}
\wick[u]{1}{[<1\psi_u(x_3)]_h[>1\psi_u(x_2)]_e}
\wick[u]{1}{<1A^{\alpha_1}(x_1)>1A^{\alpha_2}(x_2)}
\wick[u]{1}{<1A^{\alpha_3}(x_3)>1A^{\alpha_4}(x_4)} \\
& \ME{p(P')}{[\bar{\psi}_u(x_1)]_c[\bar{\psi}_u(x_3)]_g[\bar{\psi}_d(x_4)]_i}{0}
\ME{0}{[\psi_u(0)]_b[\psi_u(x_2)]_f[\psi_d(x_4)]_j}{p(P)}.
\end{split}
\nonumber
\end{equation}

In order to evaluate this expression, we have to apply the representations for the propagators and for the projection matrix elements
\pagebreak[0]
\begin{equation}
\begin{split}
& \frac{(4\pi\bar{\alpha}_s)^2e_u}{864}
\prod_{i=1}^4\int\frac{\intd^4\!x_i}{(2\pi)^4}\prod_{j=1}^2\int\frac{\intd^4\!\Delta_j}{\Delta_j^2+i0}\prod_{k=1}^2\int\frac{\intd^4\!\Lambda_k}{\Lambda_k^2+i0}
\int[\intd\!u][\intd\!v]g^{\alpha_1\alpha_2}g^{\alpha_3\alpha_4}\mathcal{S} \\
& \e^{-ix_1\cdot(\Delta_1-\Lambda_1-v_1p')}\e^{-ix_2\cdot(-\Delta_2+\Lambda_1+u_2p)}\e^{-ix_3\cdot(\Delta_2-\Lambda_2-v_2p')}\e^{-ix_4\cdot(\Lambda_2+u_3p-v_3p')}.
\end{split}
\nonumber
\end{equation}

Computing the integrations, one gets the required momentum conservation constraints
\pagebreak[0]
\begin{equation}
\begin{array}{lll}
\Delta_1 = (v_1+v_2+v_3)p'-(u_2+u_3)p & \phantom{v_2p'}\qquad & \Delta_2 = (v_2+v_3)p'-u_3p \\
\Lambda_1 = (v_2+v_3)p'-(u_2+u_3)p & & \Lambda_2 = v_3p'-u_3p.
\end{array}
\nonumber
\end{equation}

The component $\mathcal{S}$ is the sum of all required structures connected with combinations of nucleon distribution amplitudes and nucleon spinors. In order to get the desired contributions, one has to combine the twist-3 and twist-4 distribution amplitudes, studied in \cite{Braun:2000}. Furthermore, one has to specify the frame. We prefer to use the light cone decomposition given by $P_\mu=p_\mu+(m_N^2/Q^2)p'_\mu$ and $P'_\mu=p'_\mu+(m_N^2/Q^2)p_\mu$. Using this frame, we can derive the equation of motion relations and eliminate the small component of the spinor and proceed with the large component only. Let us omit the dependence on the quark momentum fractions. Moreover, we use the standard notation for the spinors.

We get the following structures for initial twist-4 and final twist-3
\pagebreak[0]
\begin{eqnarray}
\mathcal{S}_1 &=& (m_N/Q^2)\bar{N}(P')\gamma_{\alpha_4}\ppslash N(P)
\Tr [\gamma_\mu\pslash\gamma_{\alpha_2}\Deltaslash_2\gamma_{\alpha_3}\ppslash\gamma_{\alpha_1}\Deltaslash_1]
(V_1V_2+A_1A_2+V_1V_3-A_1A_3)
\nonumber \\
\mathcal{S}_2 &=& (m_N/Q^2)\bar{N}(P')\gamma_{\alpha_4}\gamma_5\ppslash N(P)
\Tr [\gamma_5\gamma_\mu\pslash\gamma_{\alpha_2}\Deltaslash_2\gamma_{\alpha_3}\ppslash\gamma_{\alpha_1}\Deltaslash_1]
(A_1V_2+V_1A_2-A_1V_3+V_1A_3)
\nonumber \\
\mathcal{S}_3 &=& (m_N/2)\bar{N}(P')\gamma_{\alpha_4}\gamma^\lambda N(P)
\Tr [\gamma_\mu\gamma_\lambda\gamma_{\alpha_2}\Deltaslash_2\gamma_{\alpha_3}\ppslash\gamma_{\alpha_1}\Deltaslash_1]
(-V_1V_3+A_1A_3)
\nonumber \\
\mathcal{S}_4 &=& (m_N/2)\bar{N}(P')\gamma_{\alpha_4}\gamma_5\gamma^\lambda N(P)
\Tr [\gamma_5\gamma_\mu\gamma_\lambda\gamma_{\alpha_2}\Deltaslash_2\gamma_{\alpha_3}\ppslash\gamma_{\alpha_1}\Deltaslash_1]
(-A_1V_3+V_1A_3)
\nonumber \\
\mathcal{S}_5 &=& (m_N)\bar{N}(P')\gamma^{\lambda'}\gamma_{\alpha_4}N(P)
\Tr [\gamma_\mu\gamma_{\alpha_2}\Deltaslash_2\gamma_{\alpha_3}i\sigma_{\lambda' p'}\gamma_{\alpha_1}\Deltaslash_1]
(T_1S_1)
\nonumber \\
\mathcal{S}_6 &=& (m_N)\bar{N}(P')\gamma^{\lambda'}\gamma_{\alpha_4}\gamma_5N(P)
\Tr [\gamma_5\gamma_\mu\gamma_{\alpha_2}\Deltaslash_2\gamma_{\alpha_3}i\sigma_{\lambda' p'}\gamma_{\alpha_1}\Deltaslash_1]
(-T_1P_1)
\nonumber \\
\mathcal{S}_7 &=& (2m_N/Q^2)\bar{N}(P')\gamma^{\lambda'}\gamma_{\alpha_4}N(P)
\Tr [\gamma_\mu i\sigma_{p p'}\gamma_{\alpha_2}\Deltaslash_2\gamma_{\alpha_3}i\sigma_{\lambda' p'}\gamma_{\alpha_1}\Deltaslash_1]
(T_1T_2-T_1T_3+T_1T_7)
\nonumber \\
\mathcal{S}_8 &=& (m_N/Q^2)\bar{N}(P')\gamma^{\lambda'}\gamma_{\alpha_4}\gamma^\lambda\ppslash N(P)
\Tr [\gamma_\mu i\sigma_{\lambda p}\gamma_{\alpha_2}\Deltaslash_2\gamma_{\alpha_3}i\sigma_{\lambda' p'}\gamma_{\alpha_1}\Deltaslash_1]
(T_1T_2+2T_1T_7)
\nonumber \\
\mathcal{S}_9 &=& (m_N/2)\bar{N}(P')\gamma^{\lambda'}\gamma_{\alpha_4}i\sigma^{\lambda\kappa}N(P)
\Tr [\gamma_\mu i\sigma_{\lambda\kappa}\gamma_{\alpha_2}\Deltaslash_2\gamma_{\alpha_3}i\sigma_{\lambda' p'}\gamma_{\alpha_1}\Deltaslash_1]
(T_1T_7).
\nonumber
\end{eqnarray}

We get the following structures for initial twist-3 and final twist-4
\pagebreak[0]
\begin{eqnarray}
\mathcal{S}_{10} &=& (m_N/Q^2)\bar{N}(P')\pslash\gamma_{\alpha_4}N(P)
\Tr [\gamma_\mu\pslash\gamma_{\alpha_2}\Deltaslash_2\gamma_{\alpha_3}\ppslash\gamma_{\alpha_1}\Deltaslash_1]
(V_2V_1+A_2A_1+V_3V_1-A_3A_1)
\nonumber \\
\mathcal{S}_{11} &=& (m_N/Q^2)\bar{N}(P')\pslash\gamma_{\alpha_4}\gamma_5N(P)
\Tr [\gamma_5\gamma_\mu\pslash\gamma_{\alpha_2}\Deltaslash_2\gamma_{\alpha_3}\ppslash\gamma_{\alpha_1}\Deltaslash_1]
(V_2A_1+A_2V_1+V_3A_1-A_3V_1)
\nonumber \\
\mathcal{S}_{12} &=& (m_N/2)\bar{N}(P')\gamma^{\lambda'}\gamma_{\alpha_4}N(P)
\Tr [\gamma_\mu\pslash\gamma_{\alpha_2}\Deltaslash_2\gamma_{\alpha_3}\gamma_{\lambda'}\gamma_{\alpha_1}\Deltaslash_1]
(-V_3V_1+A_3A_1)
\nonumber \\
\mathcal{S}_{13} &=& (m_N/2)\bar{N}(P')\gamma^{\lambda'}\gamma_{\alpha_4}\gamma_5N(P)
\Tr [\gamma_5\gamma_\mu\pslash\gamma_{\alpha_2}\Deltaslash_2\gamma_{\alpha_3}\gamma_{\lambda'}\gamma_{\alpha_1}\Deltaslash_1]
(-V_3A_1+A_3V_1)
\nonumber \\
\mathcal{S}_{14} &=& (m_N)\bar{N}(P')\gamma_{\alpha_4}\gamma^\lambda N(P)
\Tr [\gamma_\mu i\sigma_{\lambda p}\gamma_{\alpha_2}\Deltaslash_2\gamma_{\alpha_3}\gamma_{\alpha_1}\Deltaslash_1]
(-S_1T_1)
\nonumber \\
\mathcal{S}_{15} &=& (m_N)\bar{N}(P')\gamma_{\alpha_4}\gamma_5\gamma^\lambda N(P)
\Tr [\gamma_5\gamma_\mu i\sigma_{\lambda p}\gamma_{\alpha_2}\Deltaslash_2\gamma_{\alpha_3}\gamma_{\alpha_1}\Deltaslash_1]
(P_1T_1)
\nonumber \\
\mathcal{S}_{16} &=& (2m_N/Q^2)\bar{N}(P')\gamma_{\alpha_4}\gamma^\lambda N(P)
\Tr [\gamma_\mu i\sigma_{\lambda p}\gamma_{\alpha_2}\Deltaslash_2\gamma_{\alpha_3}i\sigma_{p' p}\gamma_{\alpha_1}\Deltaslash_1]
(T_2T_1-T_3T_1+T_7T_1)
\nonumber \\
\mathcal{S}_{17} &=& (m_N/Q^2)\bar{N}(P')\pslash\gamma^{\lambda'}\gamma_{\alpha_4}\gamma^{\lambda}N(P)
\Tr [\gamma_\mu i\sigma_{\lambda p}\gamma_{\alpha_2}\Deltaslash_2\gamma_{\alpha_3}i\sigma_{\lambda' p'}\gamma_{\alpha_1}\Deltaslash_1]
(T_2T_1+2T_7T_1)
\nonumber \\
\mathcal{S}_{18} &=& (m_N/2)\bar{N}(P')i\sigma^{\lambda'\kappa'}\gamma_{\alpha_4}\gamma^\lambda N(P)
\Tr [\gamma_\mu i\sigma_{\lambda p}\gamma_{\alpha_2}\Deltaslash_2\gamma_{\alpha_3}i\sigma_{\lambda'\kappa'}\gamma_{\alpha_1}\Deltaslash_1]
(T_7T_1).
\nonumber
\end{eqnarray}

Computing every structure, one always gets the dependence on $\bar{N}(P')\bar{P}_\mu N(P)$ as predicted. One can simplify the expression by exchanging
$u \leftrightarrow v$ for contributions of initial twist-3 and final twist-4. Consequently, we obtain a representation depending on initial twist-4 and final twist-3 only.

Let us now present the result of the discussed diagram depending on the integration over the quark momentum fractions. When we compare with the separated hadronic part of (\ref{ohf}), we can extract the contribution to the desired form factor. In order to get the complete result, one also needs the contributions of the other diagrams designated by $\mathcal{C}$,
\pagebreak[0]
\begin{equation}
F_2^p(Q^2)=-\frac{(4\pi\bar{\alpha}_s)^2e_u}{108}\frac{m_N^2}{Q^6}\int\frac{[\intd\!u]}{u_3(u_2+u_3)^2}\frac{[\intd\!v]}{v_3^2(v_2+v_3)^2}\mathcal{D}+\mathcal{C}.
\end{equation}

The component $\mathcal{D}$ is the sum of the remaining twist combinations of distribution amplitudes connected with multiple quark momentum fractions
\pagebreak[0]
\begin{eqnarray}
\mathcal{D}_1 &=& [V_1V_2+A_1A_2](2(u_2+u_3)(v_2+v_3)) \nonumber \\
\mathcal{D}_2 &=& [V_1V_3-A_1A_3](v_3-(u_2+u_3)(v_2+v_3)) \nonumber \\
\mathcal{D}_3 &=& [V_1A_3-A_1V_3](v_3+(u_2+u_3)(v_2+v_3)) \nonumber \\
\mathcal{D}_4 &=& [T_1S_1-T_1P_1](+2v_3-2(u_2+u_3)v_3) \nonumber \\
\mathcal{D}_5 &=& [T_1T_3+T_1T_7](-2v_3-2(u_2+u_3)v_3). \nonumber
\end{eqnarray}

Finally, we must insert the nucleon distribution amplitudes. Unfortunately, the corresponding integration is divergent. This divergency arises from endpoint singularities. That means, the integrals get divergent when a quark has no momentum or the full momentum of the nucleon. In order to analyze the structure of the divergency, one can introduce a cutoff parameter $\Omega$. Therefore, one has to respect that in case of infinite momentum transfer the integration must go from zero to one for every quark momentum fraction. According to this, we always integrate from $\Omega/Q^2$ to $1-\Omega/Q^2$, keeping in mind that the introduced parameter has the same dimension as the momentum transfer. 

Computing the modified integration, we can extract the structure of the divergency. The general behavior does not depend on the chosen polynomial expansion of the distribution amplitudes.  Moreover, this behavior is identical for all other required diagrams as well. Consequently, we can generally express the power behavior of the helicity flip form factor depending on the cutoff parameter
\pagebreak[0]
\begin{equation}
F_2^p(Q^2)\propto Q^{-6}\ln^2(Q^2/\Omega).
\end{equation}

We derived the expected power behavior of $Q^{-6}$ and we obtained a double logarithmic divergency in the case of $\Omega \rightarrow 0$. This behavior is in agreement with \cite{Belitsky:2003}.

\section{Two Photon Exchange Approximation}

Let us begin with the discussion about an important behavior of this situation. In the one photon case, the inversion of the lepton direction delivers the same contribution to the scattering amplitude. This statement is not true in the two photon case, because the inversion of the lepton direction produces another diagram. Therefore, one has to distinguish between the box diagram and the cross diagram. The corresponding contributions to the scattering amplitude must be calculated separately. 

We start with the presentation of the box diagram. In the upper part, we see the incoming electron on the left and the outgoing electron on the right. In the lower part, we have the incoming proton on the left and the outgoing proton on the right. The required quark lines denote $u$, $u$, $d$ from top to bottom. The designations at the vertices are the corresponding coordinates and the designations at the lines are the corresponding momenta.

\begin{center}
\includegraphics[scale=1.00]{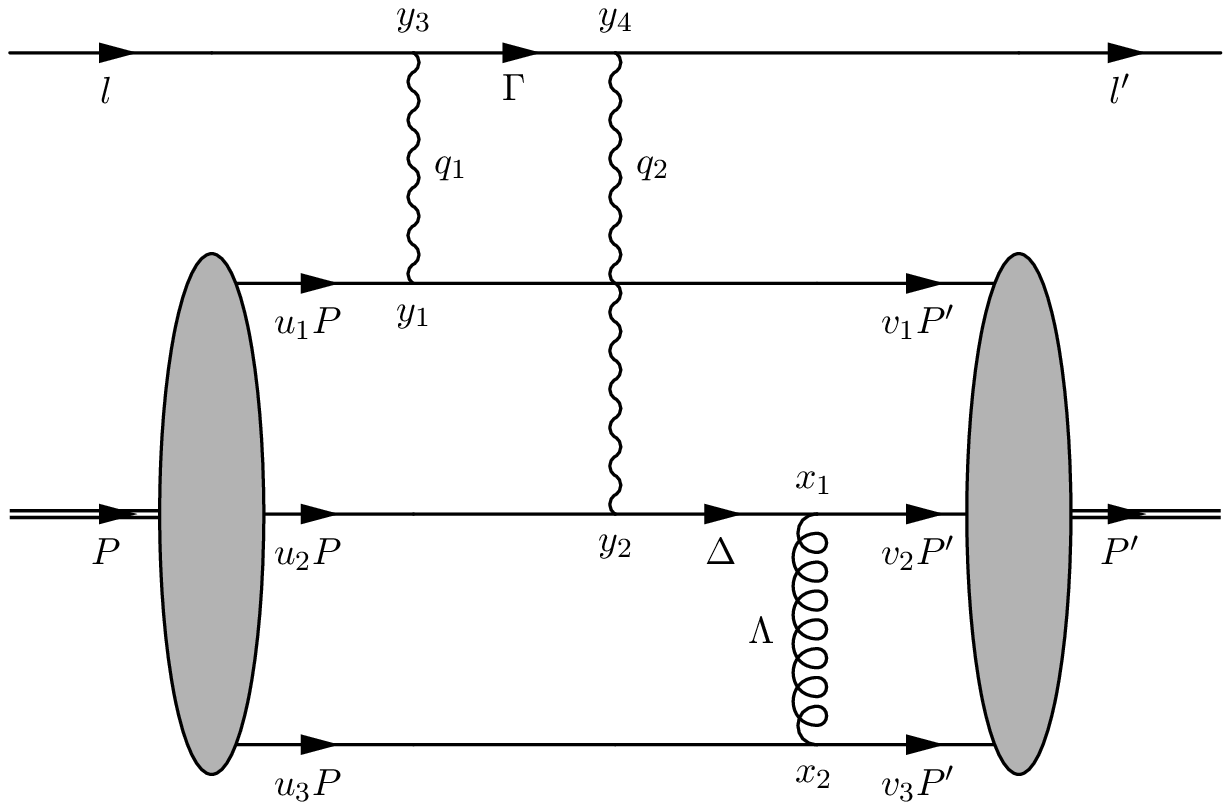}
\end{center}

We need the expression for the scattering amplitude. Applying QED Feynman rules, one can evaluate the leptonic part of the diagram directly. Moreover, one can neglect the electron mass in the propagator
\pagebreak[0]
\begin{equation}
\begin{split}
& \mathcal{M}_B=
-\frac{i(4\pi\alpha_{em})^2}{24}\prod_{i=1}^4\int\frac{\intd^4\!y_i}{(2\pi)^4}\prod_{j=1}^2\int\frac{\intd^4\!q_j}{q_j^2+i0}\int\frac{\intd^4\!\Gamma}{\Gamma^2+i0}
\bar{u}(l')\gamma^{\mu_2}\Gammaslash\gamma^{\mu_1}u(l) \\
& \ME{p(P')}{J_{\mu_2}^{em}(y_2)J_{\mu_1}^{em}(y_1)}{p(P)}\e^{iq_1\cdot(y_3-y_1)}\e^{iq_2\cdot(y_4-y_2)}\e^{-i\Gamma\cdot(y_4-y_3)}
\e^{-iy_3\cdot l}\e^{iy_4\cdot l'}.
\end{split}
\nonumber
\end{equation}

The integration over $y_3$ leads to $\Gamma=l-q_1$ and the integration over $y_4$ leads to $\Gamma=l'+q_2$. Combining them, one gets the representation $2\Gamma=(l+l')+(q_2-q_1)$. We obtain
\pagebreak[0]
\begin{equation}
\label{mb}
\begin{split}
& \mathcal{M}_B=-\frac{i(4\pi\alpha_{em})^2}{24}\prod_{i=1}^2\int\frac{\intd^4\!y_i}{(2\pi)^4}\prod_{j=1}^2\int\frac{\intd^4\!q_j}{q_j^2+i0}
\frac{1}{\Gamma^2}\bar{u}(l')\gamma^{\mu_2}\Gammaslash\gamma^{\mu_1}u(l) \\
& \ME{p(P')}{J_{\mu_2}^{em}(y_2)J_{\mu_1}^{em}(y_1)}{p(P)}\e^{-iy_1\cdot q_1}\e^{-iy_2\cdot q_2}.
\end{split}
\end{equation}

We finish with the presentation of the cross diagram. In the upper part, we see the incoming electron on the right and the outgoing electron on the left. In the lower part, we have the incoming proton on the left and the outgoing proton on the right. The required quark lines denote $u$, $u$, $d$ from top to bottom. The designations at the vertices are the corresponding coordinates and the designations at the lines are the corresponding momenta.

\begin{center}
\includegraphics[scale=1.00]{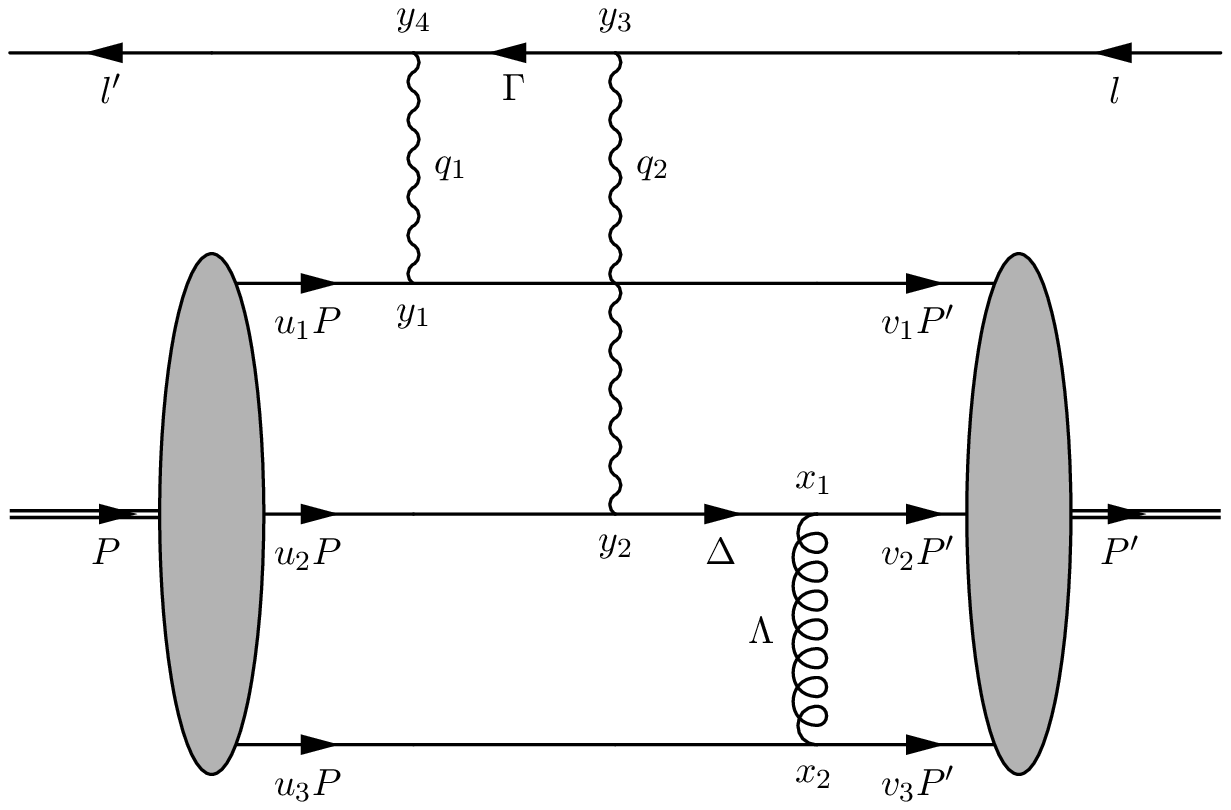}
\end{center}

We need the expression for the scattering amplitude. Applying QED Feynman rules, one can evaluate the leptonic part of the diagram directly. Furthermore, one can neglect the electron mass in the propagator
\pagebreak[0]
\begin{equation}
\begin{split}
& \mathcal{M}_C=
-\frac{i(4\pi\alpha_{em})^2}{24}\prod_{i=1}^4\int\frac{\intd^4\!y_i}{(2\pi)^4}\prod_{j=1}^2\int\frac{\intd^4\!q_j}{q_j^2+i0}\int\frac{\intd^4\!\Gamma}{\Gamma^2+i0}
\bar{u}(l')\gamma^{\mu_1}\Gammaslash\gamma^{\mu_2}u(l) \\
& \ME{p(P')}{J_{\mu_2}^{em}(y_2)J_{\mu_1}^{em}(y_1)}{p(P)}\e^{iq_1\cdot(y_4-y_1)}\e^{iq_2\cdot(y_3-y_2)}\e^{-i\Gamma\cdot(y_4-y_3)}
\e^{-iy_3\cdot l}\e^{iy_4\cdot l'}.
\end{split}
\nonumber
\end{equation}

The integration over $y_3$ leads to $\Gamma=l-q_2$ and the integration over $y_4$ leads to $\Gamma=l'+q_1$. Combining them, one gets the representation $2\Gamma=(l+l')+(q_1-q_2)$. We obtain
\pagebreak[0]
\begin{equation}
\label{mc}
\begin{split}
& \mathcal{M}_C=-\frac{i(4\pi\alpha_{em})^2}{24}\prod_{i=1}^2\int\frac{\intd^4\!y_i}{(2\pi)^4}\prod_{j=1}^2\int\frac{\intd^4\!q_j}{q_j^2+i0}
\frac{1}{\Gamma^2}\bar{u}(l')\gamma^{\mu_1}\Gammaslash\gamma^{\mu_2}u(l) \\
& \ME{p(P')}{J_{\mu_2}^{em}(y_2)J_{\mu_1}^{em}(y_1)}{p(P)}\e^{-iy_1\cdot q_1}\e^{-iy_2\cdot q_2}.
\end{split}
\end{equation}

The overall result for the scattering amplitude is given by $\mathcal{M}=\mathcal{M}_B+\mathcal{M}_C$. Let us now introduce the expansion in nucleon electromagnetic form factors. Unfortunately, the leptonic and the hadronic part are not separated in this case. Nevertheless, one can show the existence of a separated representation for $\mathcal{M}$. Whereas the basic expression just depends one the nucleon momenta, the modified expression also depends on the lepton momenta. The derivation can be taken from \cite{Guichon:2003}. Therefore, we can use $Q^2=-q^2$ and assume $q=l-l'$ together with $q=P'-P$. Furthermore, we need $2\bar{P}=P+P'$ and $2\bar{L}=l+l'$. One gets
\pagebreak[0]
\begin{equation}
\label{thf}
\mathcal{M}=\frac{i(4\pi\alpha_{em})}{Q^2}\bar{u}(l')\gamma^\mu u(l)\bar{N}(P')\left[\tilde{G}_M^p\gamma_\mu
-\tilde{F}_2^p\frac{\bar{P}_\mu}{m_N}+\tilde{F}_3^p\frac{\bar{\Lslash}\bar{P}_\mu}{m_N^2}\right]N(P).
\end{equation}

All form factors depend on $Q^2$ and one additional variable. Therefore, we choose the dimensionless quantity $\omega$ defined by $\omega=4(\bar{P}\cdot\bar{L})/Q^2$. At large $Q^2$, one gets the boundary condition $\omega \geq 1$. In principle, one can generally expand every form factor as $\tilde{F}=F+\delta F$, where $F$ is the single photon exchange contribution and $\delta F$ is the multi photon exchange contribution. We do not use this decomposition because we consider the one and two photon exchange separately. The leading form factors are considered in \cite{Kivel:2009}.

Let us now study the matrix element $\ME{p(P')}{J_{\mu_2}^{em}(y_2)J_{\mu_1}^{em}(y_1)}{p(P)}$. The evaluation of this matrix element must be combined with the other terms in (\ref{mb}) and (\ref{mc}) to derive a result for the form factor. Applying the S-matrix expansion including the interaction part of the QCD Lagrangian, one gets the following leading expression for this matrix element
\pagebreak[0]
\begin{equation}
-\frac{(4\pi\bar{\alpha}_s)}{2}\ME{p(P')}{\prod_{j=2}^1\sum_{q_j}e_{q_j}\bar{\psi}_{q_j}(y_j)\gamma_{\mu_j}\psi_{q_j}(y_j)
\T\left[\prod_{i=1}^2\int\intd^4\!x_i\sum_{q_i}\bar{\psi}_{q_i}(x_i)\gamma_{\alpha_i}A^{\alpha_i}(x_i)\psi_{q_i}(x_i)\right]}{p(P)}.
\nonumber
\end{equation}

This expansion can be described by 12 Feynman diagrams and Wick contractions. We can extract the representation of the diagram which we want to study.

Let us begin with the determination of the color factor. Therefore, one has to examine the color structure of the diagram. We denote the color indices with ($a,\ldots,f$). We get
\pagebreak[0]
\begin{equation}
\begin{split}
& \psibar{10.0} \wick[u]{1}{[<1\psi_u(x_1)]_d[>1\psi_u(y_2)]_a}
\wick[u]{1}{[t^{a_1}]_{cd}[t^{a_2}]_{ef}<1A_{\alpha_1}^{a_1}(x_1)>1A_{\alpha_2}^{a_2}(x_2)} \\
& \ME{p(P')}{[\bar{\psi}_u(y_1)]_b[\bar{\psi}_u(x_1)]_c[\bar{\psi}_d(x_2)]_e}{0}\ME{0}{[\psi_u(y_1)]_b[\psi_u(y_2)]_a[\psi_d(x_2)]_f}{p(P)}.
\end{split}
\nonumber
\end{equation}
Combining all terms and contracting the generators, one gets the color factor
\pagebreak[0]
\begin{equation}
\mathcal{C}_F = \frac{1}{6}\varepsilon_{bce}\varepsilon_{baf}\delta_{da}[t^{a_1}]_{cd}[t^{a_2}]_{ef}\delta^{a_1a_2} = -\frac{2}{3}.
\end{equation}

We continue with the evaluation of the Lorentz structure of the diagram. Therefore, we designate the Lorentz indices with ($a,\ldots,h$). Hereby, we have to distinguish between the box and cross contribution. Nevertheless, we have to include $\mathcal{C}_F$ in both representations.

Keeping in mind $2\Gamma=(l+l')+(q_2-q_1)$, one gets the expression for (\ref{mb})
\pagebreak[0]
\begin{equation}
\begin{split}
& \mathcal{M}_B=\frac{i(4\pi\bar{\alpha}_s)(4\pi\alpha_{em})^2e_u^2}{72}\prod_{i=1}^2\int\intd^4\!x_i\prod_{j=1}^2\int\frac{\intd^4\!y_j}{(2\pi)^4}
\prod_{k=1}^2\int\frac{\intd^4\!q_k}{q_k^2+i0}\frac{1}{\Gamma^2}\bar{u}(l')\gamma^{\mu_2}\Gammaslash\gamma^{\mu_1}u(l) \\
& [\gamma_{\mu_2}]_{ab}[\gamma_{\mu_1}]_{cd}[\gamma_{\alpha_1}]_{ef}[\gamma_{\alpha_2}]_{gh}
\psibar{10.2}
\wick[u]{1}{[<1\psi_u(x_1)]_f[>1\psi_u(y_2)]_a}
\wick[u]{1}{<1A^{\alpha_1}(x_1)>1A^{\alpha_2}(x_2)}
\e^{-iy_1\cdot q_1}\e^{-iy_2\cdot q_2} \\
& \ME{p(P')}{[\bar{\psi}_u(y_1)]_c[\bar{\psi}_u(x_1)]_e[\bar{\psi}_d(x_2)]_g}{0}
\ME{0}{[\psi_u(y_1)]_d[\psi_u(y_2)]_b[\psi_d(x_2)]_h}{p(P)}.
\end{split}
\nonumber
\end{equation}

Keeping in mind $2\Gamma=(l+l')+(q_1-q_2)$, one gets the expression for (\ref{mc})
\pagebreak[0]
\begin{equation}
\begin{split}
& \mathcal{M}_C=\frac{i(4\pi\bar{\alpha}_s)(4\pi\alpha_{em})^2e_u^2}{72}\prod_{i=1}^2\int\intd^4\!x_i\prod_{j=1}^2\int\frac{\intd^4\!y_j}{(2\pi)^4}
\prod_{k=1}^2\int\frac{\intd^4\!q_k}{q_k^2+i0}\frac{1}{\Gamma^2}\bar{u}(l')\gamma^{\mu_1}\Gammaslash\gamma^{\mu_2}u(l) \\
& [\gamma_{\mu_2}]_{ab}[\gamma_{\mu_1}]_{cd}[\gamma_{\alpha_1}]_{ef}[\gamma_{\alpha_2}]_{gh}
\psibar{10.2}
\wick[u]{1}{[<1\psi_u(x_1)]_f[>1\psi_u(y_2)]_a}
\wick[u]{1}{<1A^{\alpha_1}(x_1)>1A^{\alpha_2}(x_2)}
\e^{-iy_1\cdot q_1}\e^{-iy_2\cdot q_2} \\
& \ME{p(P')}{[\bar{\psi}_u(y_1)]_c[\bar{\psi}_u(x_1)]_e[\bar{\psi}_d(x_2)]_g}{0}
\ME{0}{[\psi_u(y_1)]_d[\psi_u(y_2)]_b[\psi_d(x_2)]_h}{p(P)}.
\end{split}
\nonumber
\end{equation}

In order to evaluate these expressions, we have to apply the representations for the propagators and for the projection matrix elements.

We present the expression for (\ref{mb}) with $2\Gamma=(l+l')+(q_2-q_1)$ at first
\pagebreak[0]
\begin{equation}
\begin{split}
& \mathcal{M}_B=-\frac{i(4\pi\bar{\alpha}_s)(4\pi\alpha_{em})^2e_u^2}{1152}\prod_{i=1}^2\int\frac{\intd^4\!x_i}{(2\pi)^4}\prod_{j=1}^2\int\frac{\intd^4\!y_j}{(2\pi)^4}
\prod_{k=1}^2\int\frac{\intd^4\!q_k}{q_k^2+i0}\int\frac{\intd^4\!\Delta}{\Delta^2+i0}\int\frac{\intd^4\!\Lambda}{\Lambda^2+i0} \\
& \frac{1}{\Gamma^2}\bar{u}(l')\gamma^{\mu_2}\Gammaslash\gamma^{\mu_1}u(l)\int[\intd\!u][\intd\!v]g^{\alpha_1\alpha_2}\mathcal{S} \\
& \e^{-iy_1\cdot(q_1+u_1p-v_1p')}\e^{-iy_2\cdot(q_2-\Delta+u_2p)}\e^{-ix_1\cdot(\Delta-\Lambda-v_2p')}\e^{-ix_2\cdot(\Lambda+u_3p-v_3p')}.
\end{split}
\nonumber
\end{equation}

We present the expression for (\ref{mc}) with $2\Gamma=(l+l')+(q_1-q_2)$ at last
\pagebreak[0]
\begin{equation}
\begin{split}
& \mathcal{M}_C=-\frac{i(4\pi\bar{\alpha}_s)(4\pi\alpha_{em})^2e_u^2}{1152}\prod_{i=1}^2\int\frac{\intd^4\!x_i}{(2\pi)^4}\prod_{j=1}^2\int\frac{\intd^4\!y_j}{(2\pi)^4}
\prod_{k=1}^2\int\frac{\intd^4\!q_k}{q_k^2+i0}\int\frac{\intd^4\!\Delta}{\Delta^2+i0}\int\frac{\intd^4\!\Lambda}{\Lambda^2+i0} \\
& \frac{1}{\Gamma^2}\bar{u}(l')\gamma^{\mu_1}\Gammaslash\gamma^{\mu_2}u(l)\int[\intd\!u][\intd\!v]g^{\alpha_1\alpha_2}\mathcal{S} \\
& \e^{-iy_1\cdot(q_1+u_1p-v_1p')}\e^{-iy_2\cdot(q_2-\Delta+u_2p)}\e^{-ix_1\cdot(\Delta-\Lambda-v_2p')}\e^{-ix_2\cdot(\Lambda+u_3p-v_3p')}.
\end{split}
\nonumber
\end{equation}

The appearing integrations are identical in both cases. After computation of these integrations, one gets the required momentum conservation constraints. We notice that the photon momenta do not depend on $\omega$ consequentially
\pagebreak[0]
\begin{equation}
\begin{array}{lll}
q_1 = v_1p'-u_1p & \phantom{-(u_2+u_3)p} & \Delta = (v_2+v_3)p'-u_3p \\
q_2 = (v_2+v_3)p'-(u_2+u_3)p & & \Lambda = v_3p'-u_3p.
\end{array}
\nonumber
\end{equation}

The component $\mathcal{S}$ is the sum of all required structures connected with combinations of nucleon distribution amplitudes and nucleon spinors. In order to get the desired contributions, one has to combine the twist-3 and twist-4 distribution amplitudes, studied in \cite{Braun:2000}. Furthermore, one has to specify the frame. We prefer to use the light cone decomposition given by $P_\mu=p_\mu+(m_N^2/Q^2)p'_\mu$ and $P'_\mu=p'_\mu+(m_N^2/Q^2)p_\mu$. Using this frame, we can derive the equation of motion relations and eliminate the small component of the spinor and proceed with the large component only. Let us omit the dependence on the quark momentum fractions. Moreover, we use the standard notation for the spinors.

We get the following structures for initial twist-4 and final twist-3
\pagebreak[0]
\begin{eqnarray}
\mathcal{S}_1 &=& (m_N/Q^2)\bar{N}(P')\gamma_{\alpha_2}\ppslash N(P)
\Tr [\gamma_{\mu_1}\pslash\gamma_{\mu_2}\Deltaslash\gamma_{\alpha_1}\ppslash]
(V_1V_2+A_1A_2+V_1V_3-A_1A_3)
\nonumber \\
\mathcal{S}_2 &=& (m_N/Q^2)\bar{N}(P')\gamma_{\alpha_2}\gamma_5\ppslash N(P)
\Tr [\gamma_5\gamma_{\mu_1}\pslash\gamma_{\mu_2}\Deltaslash\gamma_{\alpha_1}\ppslash]
(A_1V_2+V_1A_2-A_1V_3+V_1A_3)
\nonumber \\
\mathcal{S}_3 &=& (m_N/2)\bar{N}(P')\gamma_{\alpha_2}\gamma^\lambda N(P)
\Tr [\gamma_{\mu_1}\gamma_\lambda\gamma_{\mu_2}\Deltaslash\gamma_{\alpha_1}\ppslash]
(-V_1V_3+A_1A_3)
\nonumber \\
\mathcal{S}_4 &=& (m_N/2)\bar{N}(P')\gamma_{\alpha_2}\gamma_5\gamma^\lambda N(P)
\Tr [\gamma_5\gamma_{\mu_1}\gamma_\lambda\gamma_{\mu_2}\Deltaslash\gamma_{\alpha_1}\ppslash]
(-A_1V_3+V_1A_3)
\nonumber \\
\mathcal{S}_5 &=& (m_N)\bar{N}(P')\gamma^{\lambda'}\gamma_{\alpha_2}N(P)
\Tr [\gamma_{\mu_1}\gamma_{\mu_2}\Deltaslash\gamma_{\alpha_1}i\sigma_{\lambda' p'}]
(T_1S_1)
\nonumber \\
\mathcal{S}_6 &=& (m_N)\bar{N}(P')\gamma^{\lambda'}\gamma_{\alpha_2}\gamma_5N(P)
\Tr [\gamma_5\gamma_{\mu_1}\gamma_{\mu_2}\Deltaslash\gamma_{\alpha_1}i\sigma_{\lambda' p'}]
(-T_1P_1)
\nonumber \\
\mathcal{S}_7 &=& (2m_N/Q^2)\bar{N}(P')\gamma^{\lambda'}\gamma_{\alpha_2}N(P)
\Tr [\gamma_{\mu_1}i\sigma_{p p'}\gamma_{\mu_2}\Deltaslash\gamma_{\alpha_1}i\sigma_{\lambda' p'}]
(T_1T_2-T_1T_3+T_1T_7)
\nonumber \\
\mathcal{S}_8 &=& (m_N/Q^2)\bar{N}(P')\gamma^{\lambda'}\gamma_{\alpha_2}\gamma^\lambda\ppslash N(P)
\Tr [\gamma_{\mu_1}i\sigma_{\lambda p}\gamma_{\mu_2}\Deltaslash\gamma_{\alpha_1}i\sigma_{\lambda' p'}]
(T_1T_2+2T_1T_7)
\nonumber \\
\mathcal{S}_9 &=& (m_N/2)\bar{N}(P')\gamma^{\lambda'}\gamma_{\alpha_2}i\sigma^{\lambda\kappa}N(P)
\Tr [\gamma_{\mu_1}i\sigma_{\lambda\kappa}\gamma_{\mu_2}\Deltaslash\gamma_{\alpha_1}i\sigma_{\lambda' p'}]
(T_1T_7).
\nonumber
\end{eqnarray}

We get the following structures for initial twist-3 and final twist-4
\pagebreak[0]
\begin{eqnarray}
\mathcal{S}_{10} &=& (m_N/Q^2)\bar{N}(P')\pslash\gamma_{\alpha_2}N(P)
\Tr [\gamma_{\mu_1}\pslash\gamma_{\mu_2}\Deltaslash\gamma_{\alpha_1}\ppslash]
(V_2V_1+A_2A_1+V_3V_1-A_3A_1)
\nonumber \\
\mathcal{S}_{11} &=& (m_N/Q^2)\bar{N}(P')\pslash\gamma_{\alpha_2}\gamma_5N(P)
\Tr [\gamma_5\gamma_{\mu_1}\pslash\gamma_{\mu_2}\Deltaslash\gamma_{\alpha_1}\ppslash]
(V_2A_1+A_2V_1+V_3A_1-A_3V_1)
\nonumber \\
\mathcal{S}_{12} &=& (m_N/2)\bar{N}(P')\gamma^{\lambda'}\gamma_{\alpha_2}N(P)
\Tr [\gamma_{\mu_1}\pslash\gamma_{\mu_2}\Deltaslash\gamma_{\alpha_1}\gamma_{\lambda'}]
(-V_3V_1+A_3A_1)
\nonumber \\
\mathcal{S}_{13} &=& (m_N/2)\bar{N}(P')\gamma^{\lambda'}\gamma_{\alpha_2}\gamma_5N(P)
\Tr [\gamma_5\gamma_{\mu_1}\pslash\gamma_{\mu_2}\Deltaslash\gamma_{\alpha_1}\gamma_{\lambda'}]
(-V_3A_1+A_3V_1)
\nonumber \\
\mathcal{S}_{14} &=& (m_N)\bar{N}(P')\gamma_{\alpha_2}\gamma^\lambda N(P)
\Tr [\gamma_{\mu_1}i\sigma_{\lambda p}\gamma_{\mu_2}\Deltaslash\gamma_{\alpha_1}]
(-S_1T_1)
\nonumber \\
\mathcal{S}_{15} &=& (m_N)\bar{N}(P')\gamma_{\alpha_2}\gamma_5\gamma^\lambda N(P)
\Tr [\gamma_5\gamma_{\mu_1}i\sigma_{\lambda p}\gamma_{\mu_2}\Deltaslash\gamma_{\alpha_1}]
(P_1T_1)
\nonumber \\
\mathcal{S}_{16} &=& (2m_N/Q^2)\bar{N}(P')\gamma_{\alpha_2}\gamma^\lambda N(P)
\Tr [\gamma_{\mu_1}i\sigma_{\lambda p}\gamma_{\mu_2}\Deltaslash\gamma_{\alpha_1}i\sigma_{p' p}]
(T_2T_1-T_3T_1+T_7T_1)
\nonumber \\
\mathcal{S}_{17} &=& (m_N/Q^2)\bar{N}(P')\pslash\gamma^{\lambda'}\gamma_{\alpha_2}\gamma^{\lambda}N(P)
\Tr [\gamma_{\mu_1}i\sigma_{\lambda p}\gamma_{\mu_2}\Deltaslash\gamma_{\alpha_1}i\sigma_{\lambda' p'}]
(T_2T_1+2T_7T_1)
\nonumber \\
\mathcal{S}_{18} &=& (m_N/2)\bar{N}(P')i\sigma^{\lambda'\kappa'}\gamma_{\alpha_2}\gamma^\lambda N(P)
\Tr [\gamma_{\mu_1}i\sigma_{\lambda p}\gamma_{\mu_2}\Deltaslash\gamma_{\alpha_1}i\sigma_{\lambda'\kappa'}]
(T_7T_1).
\nonumber
\end{eqnarray}

Computing every structure, we get the dependence on multiple combinations of lepton and nucleon spinors. Nevertheless, it is possible to express all these combinations as functions of the desired component $\bar{u}(l')\gamma^\mu u(l)\bar{N}(P')\bar{P}_\mu N(P)$ only. Therefore, we have to use that the combination $\bar{u}(l')\gamma^\mu\gamma_5u(l)\bar{N}(P')\bar{P}_\mu\gamma_5N(P)$ does not contribute. Using $u_1+u_2+u_3=1$ and $v_1+v_2+v_3=1$, one gets convenient representations for all components. We notice that in the obtained result for $\mathcal{M}$, the leptonic and the hadronic part are separated now. One can apply the same twist exchange as in the previous case.

Let us now present the result of the discussed diagram connection depending on the integration over the quark momentum fractions. When we compare with (\ref{thf}), we can extract the contribution to the desired form factor. In order to get the complete result, one also needs the contributions of the other diagrams designated by $\mathcal{C}$,
\pagebreak[0]
\begin{equation}
\begin{split}
& \tilde{F}_2^p(\omega,Q^2)=\frac{(4\pi\bar{\alpha}_s)(4\pi\alpha_{em})e_u^2}{72}\frac{m_N^2}{Q^6}\int\frac{[\intd\!u]}{u_1u_3(u_2+u_3)}\frac{[\intd\!v]}{v_1v_3^2(v_2+v_3)^2} \\
& (\omega/((u_1(v_2+v_3)+(u_2+u_3)v_1)^2-(u_1-v_1)^2\omega^2))\mathcal{D}+\mathcal{C}.
\end{split}
\end{equation}

The component $\mathcal{D}$ is the sum of the remaining twist combinations of distribution amplitudes connected with multiple quark momentum fractions
\pagebreak[0]
\begin{eqnarray}
\mathcal{D}_1 &=& [V_1V_2+A_1A_2](2u_1(u_2+u_3)(v_2+v_3)+2v_1(v_2+v_3)^2) \nonumber \\
\mathcal{D}_2 &=& [V_1V_3-A_1A_3](4u_1v_3(v_2+v_3)-u_1(u_2+u_3)(v_2+v_3)-v_1(v_2+v_3)^2) \nonumber \\
\mathcal{D}_3 &=& [V_1A_3-A_1V_3](4u_1v_3(v_2+v_3)+u_1(u_2+u_3)(v_2+v_3)+v_1(v_2+v_3)^2) \nonumber \\
\mathcal{D}_4 &=& [T_1S_1-T_1P_1](+8u_1v_3(v_2+v_3)-2u_1(u_2+u_3)v_3-2v_1v_3(v_2+v_3)) \nonumber \\
\mathcal{D}_5 &=& [T_1T_3+T_1T_7](-8u_1v_3(v_2+v_3)-2u_1(u_2+u_3)v_3-2v_1v_3(v_2+v_3)). \nonumber
\end{eqnarray}

Finally, we must insert the nucleon distribution amplitudes again. Unfortunately, the corresponding integration is divergent. This behavior is similar to the one photon exchange, but now we get another singularity. This divergency just appears at the limit $\omega = 1$ and it is also an endpoint singularity. In order to analyze the structure of the divergency, we introduce an analogous cutoff parameter $\Omega$ as applied in the previous case.

Computing the modified integration, we can extract the structure of the divergency. The general behavior does not depend on the chosen polynomial expansion of the distribution amplitudes.  Moreover, this behavior is identical for all other required diagrams as well. Consequently, we can generally express the power behavior of the helicity flip form factor depending on the cutoff parameter
\pagebreak[0]
\begin{equation}
\tilde{F}_2^p(Q^2)\propto Q^{-6}\ln^2(Q^2/\Omega).
\end{equation}

We derived the expected power behavior of $Q^{-6}$ and we obtained a double logarithmic divergency in the case of $\Omega \rightarrow 0$. This behavior is in agreement with \cite{Belitsky:2003}.

\section{Conclusion and Outlook}

We want to emphasize that the result has the same power behavior for the helicity flip form factor in the one and two photon exchange approximation. According to this, one can study these contributions simultaneously. Concerning the modified helicity flip form factor of the two photon exchange, we obtained the dependence on one additional variable which can be related to the scattering angle of the experimental cross section. This behavior causes problems for the interpretation of unpolarized cross sections. Furthermore, we realize that the obtained power behavior of the helicity flip form factor can describe the experimental data based on polarized cross sections qualitatively. These conclusions can explain the different behavior in the experiments using unpolarized or polarized cross sections.

Finally, we have to discuss the required modifications to avoid the divergency. The appearing double logarithmic singularities indicate the existence of not included soft contributions. This is a consequence of the factorization approach where possible contributions from remaining soft spectator quarks are considered as power suppressed. Meanwhile, there are evidences that those contributions cannot be neglected. Using a soft effective theory, the behavior of soft contributions is discussed in \cite{Kivel:2011}. In this work, it has been pointed out that the discussed soft contributions have the same power behavior as the factorized contributions and so they must be taken into account. Unfortunately, the required techniques to get all possible soft contributions are still in development.

The studies about various nucleon form factors in multi photon exchange approximation including factorizable and non-factorizable contributions are an interesting topic which requires further investigations. Meanwhile, also comprehensive reviews were written, see \cite{Perdrisat:2007} and \cite{Arrington:2011} to get an overview about the obtained achievement.

$\phantom{X}$

We want to thank Dr. N. Kivel for useful discussions and Prof. Dr. M. V. Polyakov for enabling this work.

The work has been supported by BMBF grant 06BO9012.

\bibliographystyle{h-physrev3}
\bibliography{Literature}

\end{document}